\documentclass[12pt]{article}
\usepackage{epsf}

\begin{document}

%%%%% The following lines create the SLAC Pub Title Page
%%
\thispagestyle{empty}
\renewcommand{\thefootnote}{\fnsymbol{footnote}}

%%%%% Substitute your Pub number, month and year in the following:
%%
\begin{flushright}
{\small
SLAC--PUB--8646\\
October 2000\\}
\end{flushright}

\vspace{.8cm}

%%%%% Title and Author Information:
%%
\begin{center}
{\bf\large   
The final SLD results for $A_{LR}$ and $A_{lepton}$\footnote{Work supported by
Department of Energy contract  DE--AC03--76SF00515.}}

\vspace{1cm}

Toshinori Abe\\
Stanford Linear Accelerator Center, Stanford University,
Stanford, CA  94309\\

\medskip

Representing the SLD Collaboration
\end{center}

\vfill

\begin{center}
{\bf\large   
Abstract }
\end{center}

\begin{quote}
We present the final measurements of the left-right cross-section asymmetry 
$A_{LR}$ for $Z$ boson production by $e^+e^-$ collisions and
$Z$ boson-lepton coupling 
asymmetry parameters $A_e$, $A_{\mu}$, and $A_{\tau}$ in leptonic
$Z$ decays with the SLD detector at the SLAC Linear Collider.
Using the complete sample of polarized $Z$ bosons collected at SLD,
we get
$A_{LR}   = 0.15056\pm 0.00239$,
$A_e      = 0.1544 \pm 0.0060$, 
$A_{\mu}  = 0.142  \pm 0.015 $, and
$A_{\tau} = 0.136  \pm 0.015 $.
The $A_{LR}(\equiv A_e)$ and $A_e$ results are combined and
we find $A_e = 0.1516 \pm 0.0021$.
Assuming lepton universality, we obtain a combined effective weak mixing
angle of $\sin^2\theta^{eff}_W = 0.23098 \pm 0.00026$.
Within the context of the SM, our result prefers a light Higgs mass.
\end{quote}

\vfill

%%%%%%%%%%%%%%%
%% Choose"Presented at," "Contributed to" for conference papers
%% or "Submitted to" for journal papers
%%%%%%%%%%%%%%%
\begin{center} 
{\it Invited talk presented at} 
{\it XXXth International Conference on High Energy Physics}\\
{\it Osaka, Japan} \\
{\it July 27--August 2, 2000} \\
\end{center}

\newpage
%%
%%%%% End of title page

%%%%% Following are the commands to create the rest of the SLAC Pub.
%%
%%%%% The next two lines change the line spacing to doublespace,
%%      if you should need to do that.
%%
%\renewcommand{\baselinestretch}{2}
%\normalsize

%%%%% Your paper starts here:
%%

%% To get page numbers in the rest of the paper:
%
\pagestyle{plain}

%
% Section: Introduction
%
\section{Introduction}\label{sec:intro}%1
The SLD collaboration has reported a series of $A_{LR}$ 
measurements~\cite{cite:ALR} and
$A_e$, $A_{\mu}$, and $A_{\tau}$ measurements~\cite{cite:Alepton} 
in the production and decay of $Z$ bosons by $e^+e^-$ collisions.
$A_{LR}$ is the single best measurement of the effective weak mixing angle
($\sin^2\theta^{eff}_W$)
and has remarkably small systematic error.
The measurements of $A_e$, $A_{\mu}$, and $A_{\tau}$ improve precision for
the effective weak mixing angle measurement.
These measurements also provide a test of lepton universality and SLD makes
the only direct measurement of $A_{\mu}$.
In the context of the Standard Model (SM), these measurements provide the
best sensitivity to the Higgs mass and favor a light Higgs.
In this letter, we will present the final results of these measurements at SLD.

\section{Asymmetry measurements at SLD}\label{sec:theory}%2
Polarization-dependent differential cross section for 
$e^-_{L,R} + e^+ \to Z^0 \to f\bar{f}$ 
is expressed as follows
\[
\frac{d\sigma}{dx} \propto
 (1 - P_e A_e)(1 + x^2) + 2A_f(A_e - P_e)x
\]
where $x=\cos\theta$ is the direction of the outgoing fermion with respect to
the electron beam direction.
The signed longitudinal polarization 
of the electron beam is shown as $P_e$ 
with the convention that left-handed bunches have 
negative sign.
The asymmetry parameter is defined as 
\[
A_f = 2 v_f a_f/(v_f^2 + a_f^2)
\]
where $v_f$ and $a_f$ are the effective vector and axial-vector couplings of
the $Z$ boson to the fermion (flavor ``f'') current, respectively.
The SM assumes lepton universality and
lepton asymmetry parameters are directly related to the effective weak
mixing angle
\[
A_l \equiv \frac{2\left[1-4\sin^2\theta^{eff}_W\right]}
	{1+\left[1-4\sin^2\theta^{eff}_W\right]^2} .
\]

The polarized electron beam at SLD allows for measurements of the lepton
asymmetry parameters with two different techniques.
One is a left-right asymmetry.
The left-right asymmetry is the cross-section asymmetry for the production of 
$Z$-bosons by left-handed and right-handed electron beams.
This is sensitive to the initial state coupling ($e^+e^- \to Z$).
The other is a polarized forward-backward asymmetry.
This asymmetry is a double asymmetry which is formed by taking the difference 
in the number of forward and backward events for left-handed and right-handed
beam polarization.
This is sensitive to the final state coupling 
($Z \to e^+e^-, \mu^+\mu^-, \tau^+\tau^-$).

From hadronic final states, we measure $A_e$ using the left-right asymmetry,
which is known as the $A_{LR}$ measurement.
Leptonic final states provide $A_e$ from the left-right asymmetry and
$A_e$, $A_{\mu}$, and $A_{\tau}$ from the polarized forward-backward asymmetry.
%We combine these $A_e$, $A_{\mu}$ and $A_{\tau}$ measurements to form 
%$A_{lepton}$.
We compare these asymmetry measurements to test lepton universality.
Assuming universality, we combine $A_{LR}$ with $A_e$, $A_{\mu}$, 
and $A_{\tau}$ to derive our grand average effective weak mixing angle.

SLD has collected polarized $Z$ data from 1992 to 1998.
We collected about 530K polarized $Z$ events with about 75\% electron beam
polarization.

\section{The $A_{LR}$ measurement}\label{sec:ALR}%4
The event selection for the $A_{LR}$ measurement requires 
a hadronic signature and discriminates against beam background,
two photon, and $e^+e^-$ final states events.
The selection efficiency is about 91\% for hadronic final states.
There is a small amount of $\tau^+\tau^-$ final state (0.3\%) 
which is not background.
The background fraction is only 0.04\%.
Using selected events, we extract the measured cross-section asymmetry 
$A_m$ as follows
\[
A_m = \frac{N_L - N_R}{N_L + N_R}
\]
where $N_L$ ($N_R$) is a number of selected events for left-handed
(right-handed) electron beam.
We need two more steps to evaluate the result.
First, we correct background and small machine/beam related asymmetries,
which are small, and divide by the measured polarization as follows,
\[
 A_{LR} = \frac{A_m}{P_e} +\frac{1}{P_e}\delta A_m=\frac{A_m}{P_e}+O(10^{-4}) .
\]
Next, $A_{LR}$ is converted to the $Z$-pole result by applying $\gamma Z$
interference and initial state radiation corrections
\[
 A_{LR}^0 = A_{LR}+\delta A_{EW} .
\]
This calculation requires knowledge of our luminosity weighted mean 
center-of-mass energy.
The relative size of the correction is about 2\%.

\begin{figure}[h]%1
\centering
\epsfxsize=3.1in
\epsffile{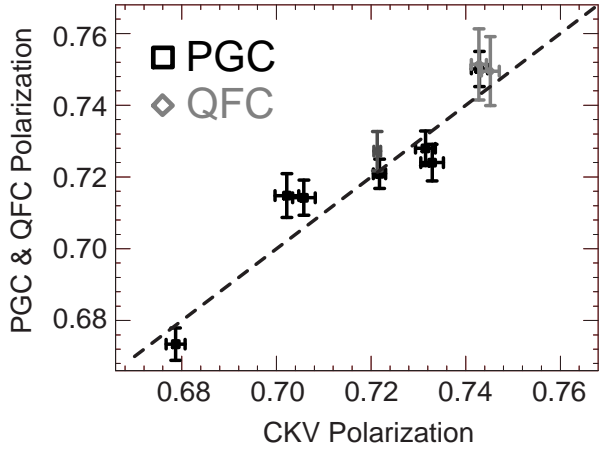}
\caption{Comparison of PGC and QFC polarizations measurements 
to the one from the primary polarimeter CKV.}
\label{fig:pol}
\end{figure}

The electron polarization plays an important role in the measurement.
There are three detectors to measure the electron polarization.
Our primary polarimeter is the Cherenov detector (CKV), which detects 
Compton-scattered electrons.
Polarized Gamma Counter (PGC) and Quartz Fiber Calorimeter (QFC) are
used to assist in the calibration of CKV and detect Compton-scattered
photons.
Fig.~\ref{fig:pol} shows a comparison of measured polarizations
with these detectors.
The measurements of electron polarization by these detectors are
consistent and the systematic error from the calibration uncertainty is
0.40\%.
The obtained total systematic uncertainty from polarization measurement 
is 0.50\%, which is the biggest
systematic error source for the $A_{LR}$ measurement.

We performed two additional systematic error checks in the 1997-98
run, of the energy scale and positron polarization.
For precise understanding of the average center-of-mass energy, we did a
$Z$-pole scan.
We measured two off-peak data points and obtained the average
center-of-mass energy of $\sqrt{s}=91.237\pm0.029$ GeV.
The uncertainty of the center-of-mass energy leads to systematic error
of 0.39\% on $A_{LR}^0$ due to energy dependence of the $\gamma Z$
interference and initial state radiation corrections.
This is the second biggest systematic error source of the $A_{LR}$
measurement.
In the past, we had assumed positron polarization is zero.
We now have directly measured the positron polarization with a M\o ller
polarimeter in End Station A and obtained a result of 
($-0.02 \pm 0.07$)\%
which is consistent with zero.

The final result of the $A_{LR}$ measurement is 
\[
\begin{array}{lclcl}
A_{LR}^0 & = & 0.15138 & \pm & 0.00216 ,\\
\sin^2\theta^{eff}_W & = & 0.23097 & \pm & 0.00027 .
\end{array}
\]
The systematic error in the effective weak mixing angle measurement
is about 0.0001.
Our error is still dominated by the statistical error.

\section{$A_e$, $A_{\mu}$, and $A_{\tau}$ measurements}\label{sec:Al}%4
\begin{figure}[h]%2
\centering
\epsfxsize=3.1in
\epsffile{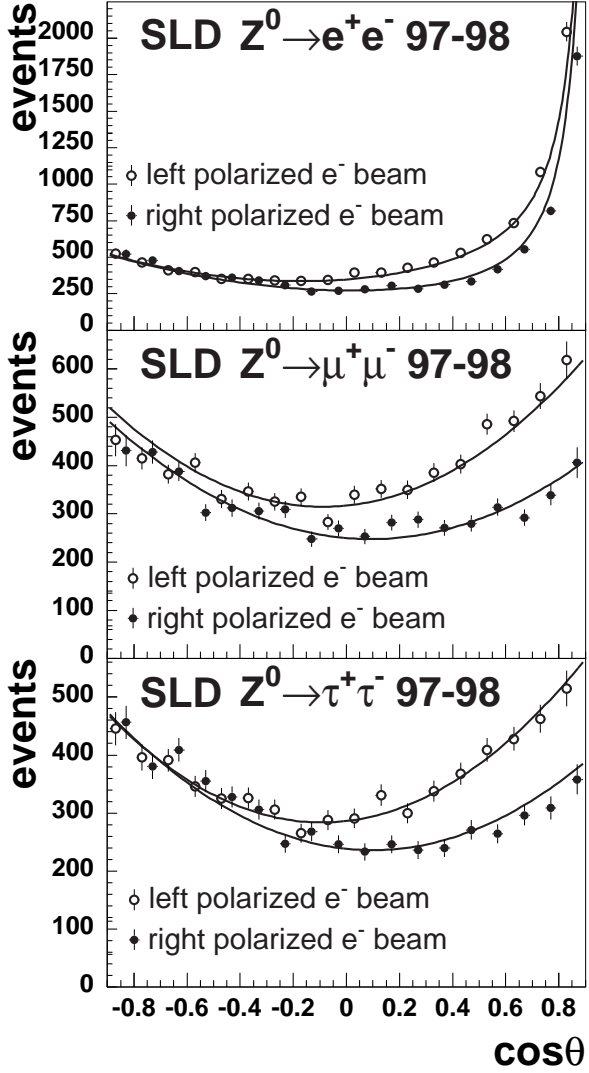}
\caption{Polar-angle distributions for $Z$ decays to $e$, $\mu$, 
and $\tau$ pairs.}
\label{fig:lepton}
\end{figure}

Leptonic $Z$ decay candidates are required to have low charged multiplicity and
two back-to-back leptons (or in the case of the tau-pair events, the tau decay
products).
The selection efficiencies for each lepton species are $70\sim75$\% and
their purities are about 99\% except tau (about 95\%).
Fig.~\ref{fig:lepton} shows angular distributions of selected leptonic final 
states for left- and right-handed electrons
(Selection efficiency is corrected in the figure).

Using these events, $A_e$ and $A_{\mu}$ or $A_{\tau}$ are simultaneously
determined from an unbinned maximum liklihood functions include $Z$,
$\gamma Z$, and $\gamma$ cross-section terms, 
and initial state radiation effects.
We find the results for $A_e$, $A_\mu$, and $A_\tau$ from leptonic $Z$
decay events are
\[
\begin{array}{lclcll}
A_e    & = & 0.1544 & \pm & 0.0060 &,\\
A_\mu  & = & 0.142 & \pm & 0.015  &\mbox{, and}\\
A_\tau & = & 0.136 & \pm & 0.015 .&
\end{array}
\]

\section{The SLD grand average result}\label{sec:SLDav}%4
The $A_{LR}$ measurement measures the initial state
coupling ($A_{LR}^0 \equiv A_e$).
Hence we combine the $A_{LR}^0$ result and 
$A_e$ from purely leptonic final states
taking care of small effects due to correlations in systematic uncertainties,
and obtain 
\[
A_{LR}^0+A_e=0.1516 \pm 0.0021 .
\]
Our results are consistent with lepton universality.
Therefore we can assume universality and the obtained
our grand average result of the lepton asymmetry parameter 
and the effective weak mixing angle are
\[
\begin{array}{lclcl}
A_{l} & = & 0.15130 & \pm & 0.00207 ,\\
\sin^2\theta^{eff}_W & = & 0.23098 & \pm & 0.00026 .
\end{array}
\]

\section{World effective weak mixing angle measurements}
\begin{figure}[h]%3
\centering
\epsfxsize=3.5in
\epsffile{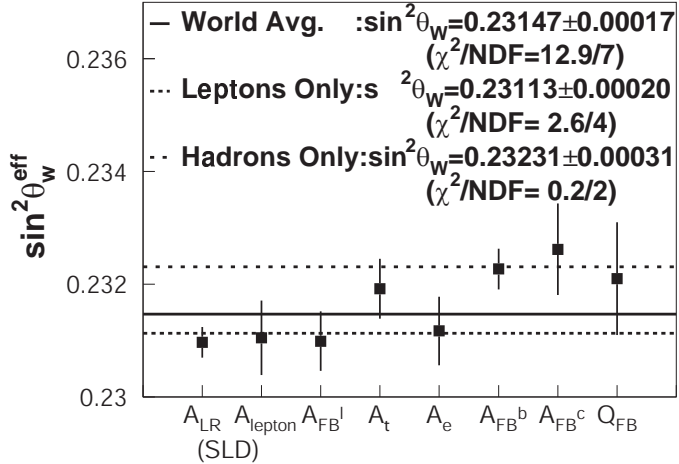}
\caption{The world effective weak mixing angle measurements.}
\label{fig:sstw}
\end{figure}
Now we compare our result with other measurements~\cite{:2000nr}.
Fig.~\ref{fig:sstw} shows the world effective weak mixing angle measurements.
The world average value is $\sin^2\theta^{eff}_W = 0.23147\pm0.00017$.
Results with leptonic asymmetry are consistent each other 
($\chi^2/NDF=2.6/4$)
and the
average value is $\sin^2\theta^{eff}_W = 0.23113\pm0.00020$.
Results with hadronic technique ($\sin^2\theta^{eff}_W =
0.23231\pm0.00031$) are self-consistent ($\chi^2/NDF=0.2/2$).
However there is 3$\sigma$ difference between leptons only and hadrons
only results.

\begin{figure}[h]%4
\centering
\epsfxsize=3.1in
\epsffile{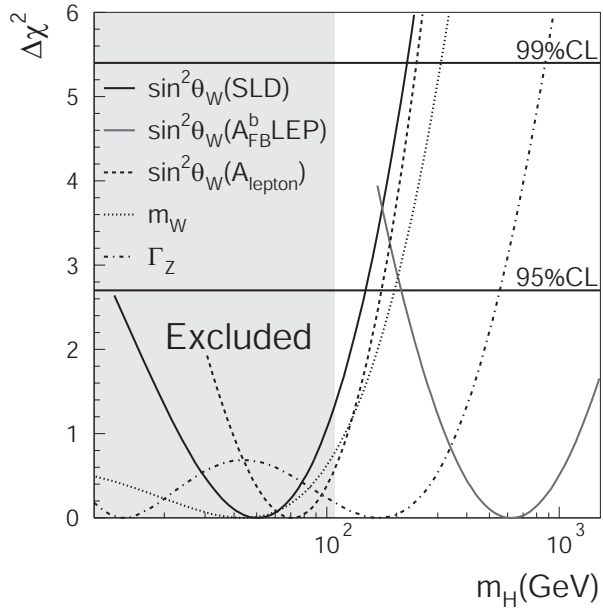}
\caption{Higgs mass plot by technique.}
\label{fig:Higgs}
\end{figure}

Since the effective weak mixing angle is very sensitive to the Higgs mass,
it is interesting to derive the allowed Higgs mass region.
We use the measured $Z$ boson~\cite{:2000nr} and 
top quark~\cite{Cite:Top} masses, a determination of 
$\alpha(M_Z^2)$~\cite{Martin:2000by},
and the ZFITTER~6.23 program~\cite{Bardin:1999yd}
to obtain the results.
Fig.~\ref{fig:Higgs} shows the allowed Higgs mass regions given by the
individual measurements.
The SLD result prefers a light Higgs mass~\cite{Cite:Higgs}.
On the other hand, 
the result by the hadron technique expects a heavy Higgs mass.
There are several measurements sensitive to the Higgs mass,
W mass ($m_W$) and $Z$ width ($\Gamma_Z$) measurements.
The allowed Higgs mass regions by the measurements are also shown in 
Fig.~\ref{fig:Higgs}.
The most sensitive measurement to Higgs mass is given by the effective
weak mixing angle by the lepton technique.
The curve given by W mass measurement and the result is in good
agreement with the one by the lepton technique.
These results prefer a light Higgs mass.

\section{Conclusions}
The SLD collaboration has finalized its very precise measurement of the
weak mixing angle and the result is $\sin^2\theta^{eff}_W=0.23098\pm0.00026$.
The SLD error is equivalent to an W mass measurement uncertainty of 38MeV 
assuming the SM, 
which is equal to the error of global direct $m_W$ measurements
from FERMILAB and LEP II.
The SLD/LEP lepton asymmetry measurements are self-consistent.
In the context of the SM, our data prefers a light Higgs mass.

%%%%% Acknowledgments
%%
%\subsection*{Acknowledgments}
%
%Notice that the Acknowledgments section is unnumbered.
%We thank A. Physicist and B. Physicist for discussions.
%

%%%%% Bibliography
%%
 
%%
%%%%% End Bibliography

\end{document}